%% file: 1bsmhiggs.tex
\documentclass[10pt]{article}
\usepackage{graphicx}

\def\Title#1{\begin{center} {\Large #1 } \end{center}}
\def\Author#1{\begin{center}{ \sc #1} \end{center}}
\def\Address#1{\begin{center}{ \it #1} \end{center}}

\newcommand\pubblock{\rightline{\begin{tabular}{l} Proceedings of the Second Annual LHCP\\ \pubnumber\\
         \pubdate  \end{tabular}}}

\newenvironment{Abstract}{\begin{quotation} \begin{center} 
             \large ABSTRACT \end{center}\bigskip 
      \begin{center}\begin{large}}{\end{large}\end{center} \end{quotation}}

\newenvironment{Presented}{\begin{quotation} \begin{center} 
             PRESENTED AT\end{center}\bigskip 
      \begin{center}\begin{large}}{\end{large}\end{center} \end{quotation}}

\def\Acknowledgements{\bigskip  \bigskip \begin{center} \begin{large}
             \bf ACKNOWLEDGEMENTS \end{large}\end{center}}

\input econfmacros.tex

\usepackage{color}

\newcommand\pubnumber{ }

\newcommand\pubdate{}

\def\affiliation{ }

\begin{document}

\large
\begin{titlepage}
\pubblock

\vfill
\Title{Higgs Physics Beyond the Standard Model}
\vfill

\Author{ Margarete Muhlleitner  }
\Address{\affiliation Institute for Theoretical Physics, Karlsruhe
  Institute of Technology, 76128 Karlsruhe, Germany}
\vfill
\begin{Abstract}
\noindent 
Higgs physics beyond the Standard Model (SM) is presented in the
context of an underlying strong dynamics of electroweak symmetry
breaking (EWSB) as given by composite Higgs models. Subsequently, the
study of New Physics (NP) effects in a more model-independent way 
through the effective Lagrangian approach is briefly sketched before moving on to
the investigation of NP through Higgs coupling measurements. Depending
on the precision on the extracted couplings, NP scales up to the TeV
range can be probed at the high-luminosity option of the LHC, if
the coupling deviations arise from mixing effects or from some underlying
strong dynamics.
\end{Abstract}
\vfill

\begin{Presented}
The Second Annual Conference\\
 on Large Hadron Collider Physics \\
Columbia University, New York, U.S.A \\ 
June 2-7, 2014
\end{Presented}
\vfill
\end{titlepage}
\def\thefootnote{\fnsymbol{footnote}}
\setcounter{footnote}{0}
%

\normalsize 


\section{Introduction}
The announcement of the discovery of a new scalar particle by the
Large Hadron Collider (LHC) experiments ATLAS and CMS
\cite{:2012gk,:2012gu} has immediately triggered activities to determine
the properties of this particle. The measurement of its couplings to
other SM particles and the extraction of its spin and parity quantum 
numbers are steps in order to establish the scalar as {\it
  the} Higgs boson, {\it i.e.}~the particle related to EWSB. Any
deviations in these properties from the 
SM expectation would hint to physics beyond the SM
(BSM). Although the Higgs boson looks very SM-like, there is still
room for interpretations within BSM theories. In absence of any direct
detection of NP particles, the Higgs sector
becomes particularly interesting. The precise measurements
of the Higgs properties help to reveal the underlying mechanism of
EWSB and in particular may shed light on the question if the
underlying dynamics is strongly or weakly interacting. An example for
the latter are supersymmetric (SUSY) extensions of the SM which remain
weakly interacting up to high energies. The talk, that is summarised
here, focuses on non-SUSY extensions of the SM. Composite
Higgs Models shall be presented as theories emerging from a strongly
interacting sector, before moving on to effective theory descriptions
that allow for a more model-independent investigation of the Higgs
sector. The last part finally is dedicated to specific (non-SUSY) models and
how they can be probed through coupling measurements. 
\begin{figure}[!b]
\centering
\includegraphics[scale=0.3, angle=-90]{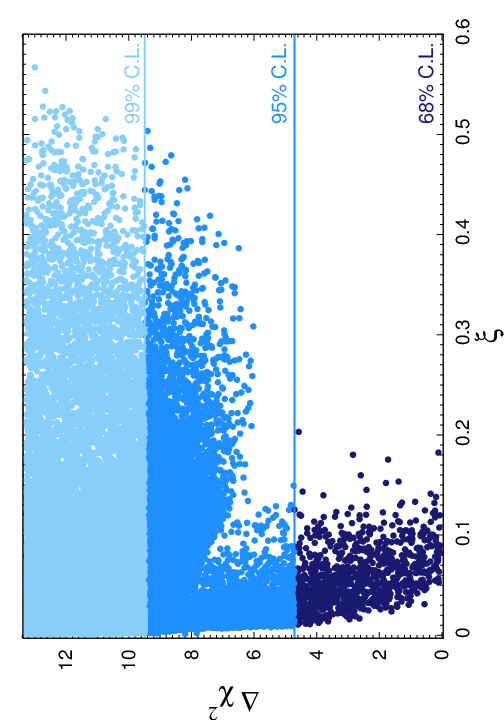}
\caption{Parameters passing the $\chi^2$ test of electroweak precision
 observables, which fulfill $|V_{tb}| > 0.92$, \cite{bquarkcomp}.}
\label{fig:ewptcomp}
\end{figure}
%

\section{Composite Higgs Boson}
In composite Higgs models the Higgs boson arises as a pseudo
Nambu-Goldstone boson from a strongly-coupled sector \cite{strong}. As
result of the Goldstone nature of the Higgs boson, in the Strongly
Interacting Light Higgs (SILH) scenario \cite{silh} there is a light narrow
Higgs-like scalar, which is the bound state from some strong dynamics
and which is separated by a mass gap from the other usual resonances of the
strong dynamics. At low energy, the particle content is hence the same
as in the SM. The Higgs couplings to the SM particles, however,
are modified \cite{silh}. In composite Higgs models the problem of the generation
of fermion masses is solved by the hypothesis of partial
compositeness \cite{partial}. The SM fermions, which are elementary,
couple linearly to heavy states of the strong sector with the same
quantum numbers, implying in particular the top quark to be largely
composite. The global symmetry of the strong sector is explicitly
broken by these couplings, and the Higgs potential is generated from
loops of SM particles with the dominant contribution coming from
the top quark. 
As has been shown in Ref.~\cite{lowmass} a low-mass Higgs
boson of $\sim 125$~GeV can naturally be accommodated only if the heavy
quark partners are rather light, with masses below about $1$~TeV. From a
phenomenological point of view, the modified Higgs couplings to the SM
particles not only change the Higgs production and decay rates
\cite{rates,ehdecayprogram}, but notably lead to an increase of the
cross section for double Higgs production in vector boson fusion with the
energy \cite{silh,strongdouble}. 
Furthermore, composite Higgs models are
challenged by electroweak precision tests (EWPTs) \cite{silh,ewpt}. The
tension with the $S$ and $T$ parameters \cite{peskin} can be weakened 
through the contributions from new heavy fermions
\cite{wfermions,bquarkcomp}. This is shown in Fig.~\ref{fig:ewptcomp} for a
model with composite bottom
quarks, where the fermions are embedded in the {\bf 10}, the smallest
possible representation of $SO(5)$ that allows for partially composite
bottom quarks while being compatible with the EWPTs by implementing
custodial symmetry. Performing a $\chi^2$ test, taking into account
the EWPTs and the measurement of $V_{tb}$ \cite{ckm}, it displays $\Delta
\chi^2 = \chi^2 - \chi^2_{\scriptsize \mbox{min}}$ as a function of
$\xi$ for the points passing the test after a scan over the model parameters. Here
$\xi = v^2/f^2$, where $v \approx 246$~GeV  is the vacuum expectation
value and $f$ the typical scale of the Goldstone bosons of the strong sector.

\begin{figure}[h]
\centering
\includegraphics[scale=0.55]{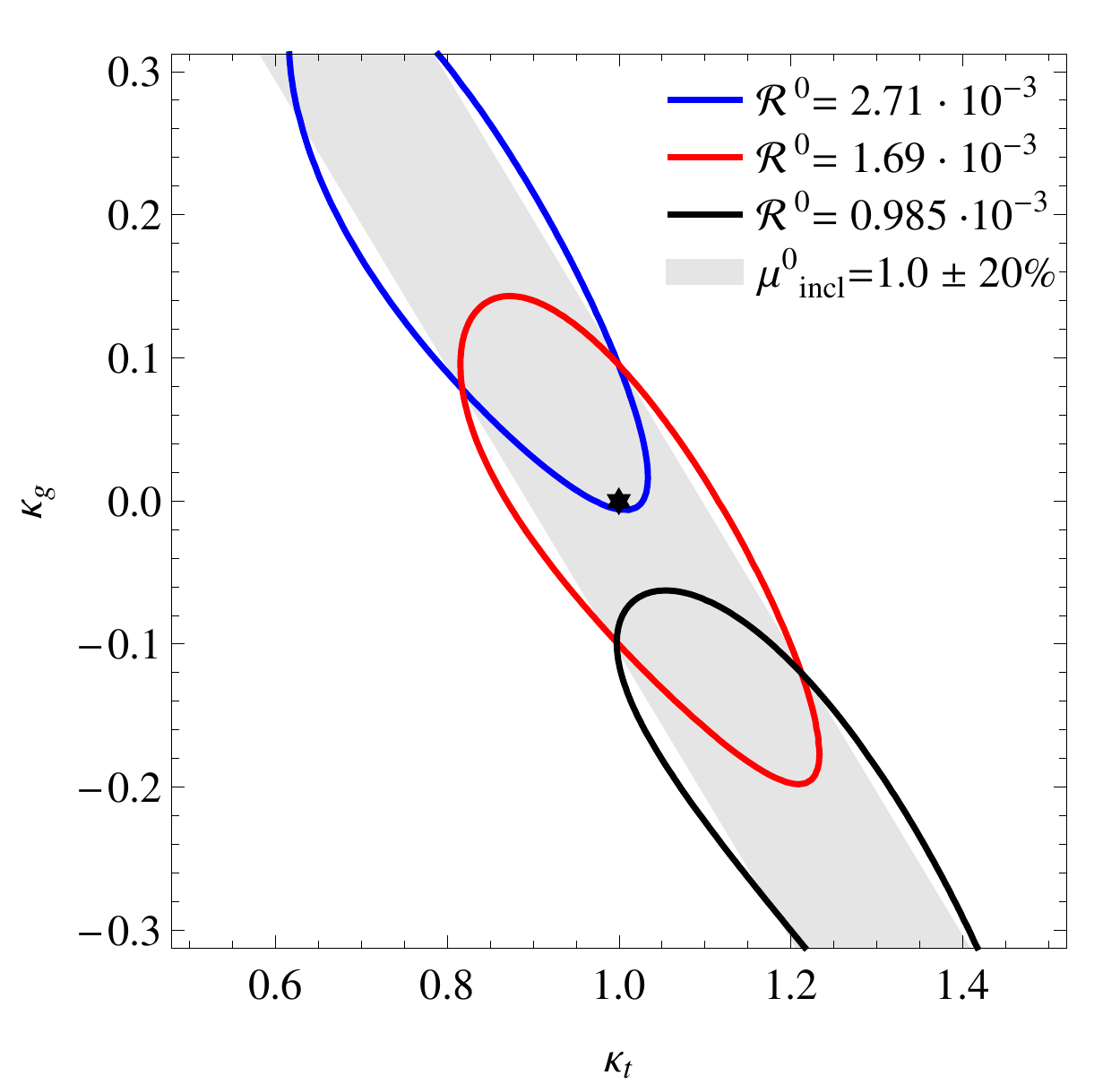}
\caption{The 95\% confidence level (C.L.) contours from a $\chi^2$
  test, in the plane of the top and effective gluon coupling
  modifiers $\kappa_t$ and $\kappa_g$ for different values of
  the inclusive signal strength $\mu^0_{\scriptsize \mathrm{incl}}$ and the boosted
  observable ${\cal R}^0$. For details, see \cite{schlaffereal}.}
\label{fig:resolvecoup}
\end{figure}

As long as no heavy fermion partners have been detected directly,
their influence on loop induced processes like Higgs production
through gluon fusion becomes particularly interesting. It has been
shown, that the process computed by applying the Low-Energy Theorem
(LET) \cite{let} is insensitive to the details of the couplings and masses of the
strong sector \cite{singlehindep,letappl}. In double Higgs
production, however, the LET is not reliable any more \cite{baur} and the
cross section becomes sensitive to the properties of the strong sector
\cite{letappl,anomalous}. Also the production of a boosted Higgs boson
in association with a high-transverse momentum jet is sensitive to the
details of heavy fermions
\cite{higgspjets,schlaffereal} and can be exploited 
to measure the Higgs coupling to top pairs, as shown in
Fig.~\ref{fig:resolvecoup} from Ref.~\cite{schlaffereal}.

\section{Effective Lagrangian Approach}
The plethora of NP extensions calls for an effective framework that
captures NP effects in a model-independent way. The scale $\Lambda$ at
which NP becomes effective, proposed by natural mechanisms of EWSB, is
not far from the TeV scale, so that a convenient framework for a
model-independent analysis of deviations from the SM is given by an
effective Lagrangian approach which enlarges the SM by including
higher-dimensional operators built from SM fields
\cite{dimsix1,dimsix2}. At the dimension-6 level, there are 59
linearly independent operators, taking into account only one fermion
generation and allowing also for CP-odd operators 
\cite{dimsix2,ehdecay1}. There has been some discussion not only about
the most appropriate basis to use but also about the minimum number of
operators that should be applied to best capture the NP impact on
Higgs physics. As it is impossible to review in this short text all 
contributions in this field a few recent results shall be highlighted
in the following. The authors of Ref.~\cite{Elias-Miro:2013mua} found that for one
family there are 8 CP-even operators that, at tree-level, can only affect Higgs
physics and no other SM processes. In a bottom-up approach, taking as
starting point all possible new interactions among SM fields, the
authors of Refs.~\cite{primaries,pomarolriva} derived the set of independent new
interactions at the dimension-6 level, that are presently best tested
by the experiments and that give the best way to constrain NP. They found
18 of these BSM primary effects, not taking into 
account four-fermion deviations, minimal flavour violation suppressed
deviations and those arising from CP violation. There are 8 Higgs
primaries, 7 EWPT primaries and 3 primaries affecting triple gauge
boson couplings (TGC). All other NP effects are not independent and
are correlated with the BSM primaries, see also
\cite{ehdecay1,Brivio:2013pma}. In particular it is found, that large
NP effects can still be revealed in the Higgs decay $H \to Z
\gamma$ \cite{Dumont:2013wma}, while the deviations from the SM in the
differential distributions of Higgs decays into a vector boson and a fermion pair
are already constrained from TGC measurements \cite{pomarolriva}.

\section{Higgs Coupling Measurements as a Probe for New 
Physics}
The effective field theory approach has the advantage to allow for the
study of a large class of models in a rather model-independent
way. However, it cannot account for effects from light particles in
the loops or for Higgs decays into light non-SM particles. In order to
give a complete picture of BSM effects in the Higgs sector the
effective approach therefore has to be complemented by studies in
specific models, that ideally capture these features. 
The subject of this section is the information that can be obtained from the
measurement of the Higgs couplings, in particular also on the scale of NP,
both in the effective Lagrangian approach and in specific 
models. For a review, see \cite{coupreview}. Deviations in the 
Higgs couplings due to NP can occur from two effects. The couplings can
be modified due to the mixing of the standard Higgs
field with other scalar fields. This is {\it e.g.}~the case in portal models, where the
SM Higgs field is coupled with a hidden sector, or in extensions of the
simplest Higgs sector by a second Higgs doublet. The second class of
mixing effects arises from vertex corrections of Higgs couplings to SM
particles due to virtual contributions of new gauge bosons, scalars or
fermions. Such loop effects can occur in various models like {\it
  e.g.}~supersymmetry, extra dimensions, see-saw models, strong
dynamics or extended gauge groups. 

Characterising NP effects by higher dimensional operators
\cite{dimsix1,dimsix2}, the deviations of the Higgs couplings $g$ from
the corresponding SM couplings $g_{\scriptsize \mbox{SM}}$ are of the
order of
\begin{eqnarray}
g = g_{\scriptsize \mbox{SM}} [1+\Delta] \; , \quad \mbox{ with }
\Delta = {\cal O} (v^2/\Lambda^2) \;, \label{eq:coupdev}
\end{eqnarray}
where $\Lambda \gg v$ is the characteristic BSM
scale.\footnote{This does not hold in case the underlying model
violates the decoupling theorem \cite{Appelquist:1974tg}.} Depending
on the precision $\Delta$ with which the couplings are measured this
allows then to probe mass scales of the order of $\Lambda =
\frac{v}{\sqrt{\Delta}} $.
For coupling modifications that are generated by loop effects, there is
an additional loop suppression factor $1/(16 \pi^2)$ that adds to
potentially small couplings between the SM and the new particles, so
that only scales up to $\Lambda < v/(4\pi \sqrt{\Delta})$ can be
probed. Loop effects are therefore less promising for the indirect
exploration of NP scales than mixing effects.

\begin{figure}[h]
\begin{center}
\includegraphics[width=0.6\textwidth]{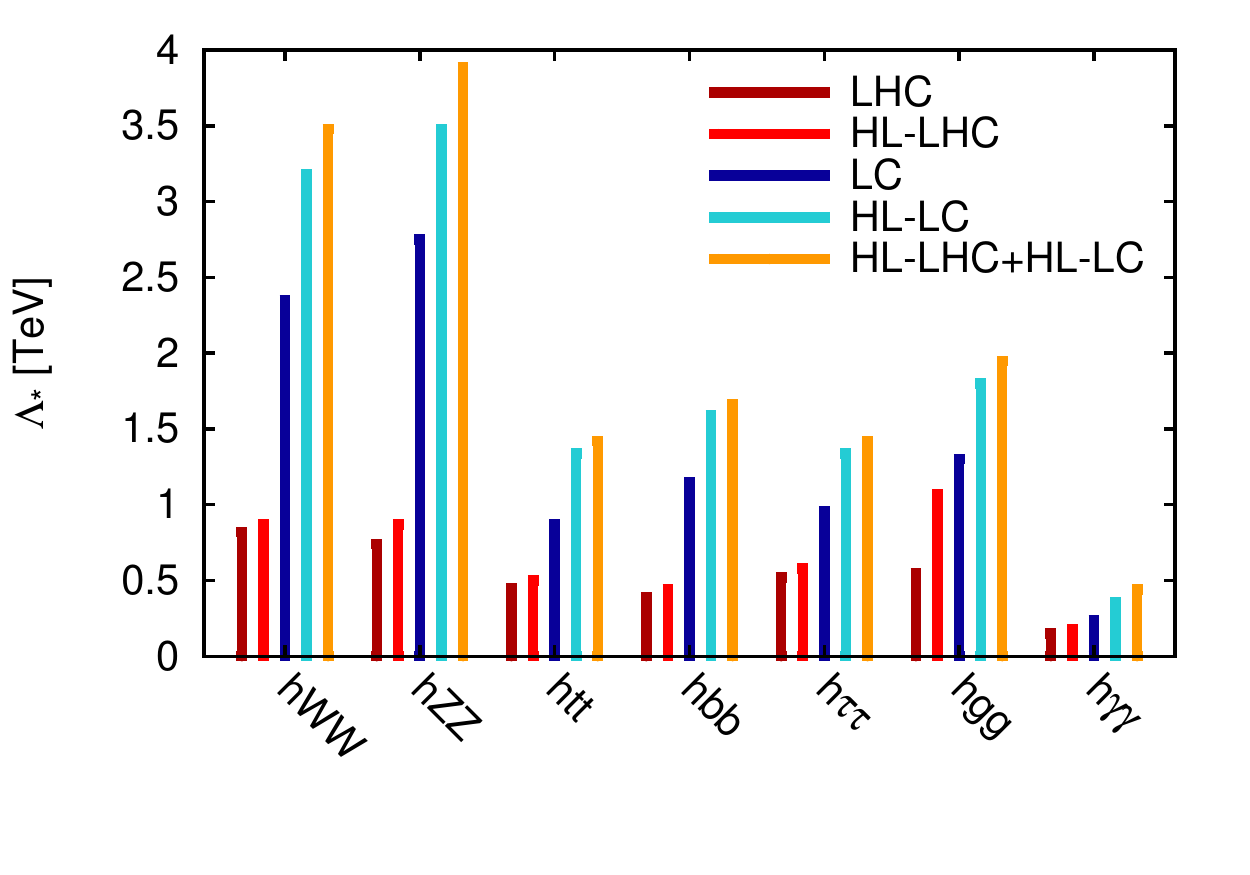}
\vspace{-1.2cm}
\caption{Effective NP scales $\Lambda_\ast$ extracted from the
	 Higgs coupling measurements collected in
	 Table~\ref{tab:coupprec}. (The ordering of
	 the columns from left to right corresponds to the legend
	 from up to down.) For details, see \cite{coupreview}.}
\label{fig:limscales}
\end{center}
\end{figure}

Table~\ref{tab:coupprec} summarises the present precision on the
couplings from measurements at the LHC \cite{presentlhc,both} and the
accuracy that can be achieved at the high-luminosity (HL) run of the LHC,
at a future $e^+e^-$ linear collider (LC) \cite{both,lincol} and from
the combination of the HL-LHC and HL-LC results. The extracted limits
on the effective scales $\Lambda_*$ from the contributions of the
dimension-6 operators taking into 
account these coupling precisions are shown in
Fig.~\ref{fig:limscales}. 
\begin{table}[t!]
\begin{center}
\begin{tabular}{|l||c|c||c|c||c|}
\hline
coupling	    &  LHC  & HL-LHC	       & LC	   & HL-LC & HL-LHC + HL-LC \\
\hline\hline
$hWW$		    &  0.09 & 0.08	       & 0.011	   & 0.006 & 0.005	    \\
$hZZ$		    &  0.11 & 0.08	       & 0.008	   & 0.005 & 0.004	    \\
$htt$		    &  0.15 & 0.12	       & 0.040	   & 0.017 & 0.015	    \\
$hbb$		    &  0.20 & 0.16	       & 0.023	   & 0.012 & 0.011	    \\
$h\tau\tau$	    &  0.11 & 0.09	       & 0.033	   & 0.017 & 0.015	    \\
\hline
$h\gamma\gamma$     &  0.20 & 0.15	       & 0.083	   & 0.035 & 0.024	    \\
$hgg$		    &  0.30 & 0.08	       & 0.054	   & 0.028 & 0.024	    \\
\hline
$h_{\scriptsize \mathrm{invis}}$   &	---  & ---		& 0.008     & 0.004 & 0.004	     \\
\hline
\end{tabular}
\vspace*{0.2cm}
\caption{Expected accuracy at the 68\% C.L.\ with which fundamental and derived
  Higgs couplings can be measured; the deviations are defined as
  $g=g_{\scriptsize \mbox{SM}} [1\pm\Delta]$ compared to the Standard Model at the LHC/HL-LHC (luminosities
  300 and 3000 fb$^{-1}$), LC/HL-LC (energies 250+500~GeV / 250+500~GeV+1~TeV and luminosities
  250+500 fb$^{-1}$ / 1150+1600+2500 fb$^{-1}$), and in combined analyses of HL-LHC and HL-LC.
  For invisible Higgs decays the upper limit on the underlying
  couplings is given. Taken from \cite{coupreview}.}
\label{tab:coupprec}
\vspace*{-0.5cm}
\end{center}
\end{table}
They have been obtained with SFitter
\cite{presentlhc} after defining the effective scales $\Lambda_*$, that are
obtained by factoring out from the operators typical coefficients like
couplings and loop factors. Furthermore, in the loop-induced couplings
to the gluons and photons only the contributions from the contact
terms are kept. The effects of the loop terms are already disentangled
at the level of the input values $\Delta$. 

In the context of composite Higgs models, based on the estimates of
potential deviations from SM Higgs couplings, bounds have been derived on
the compositeness parameter $\xi$. These in turn translate into bounds
on the compositeness scale $f$ and are summarised in
Table~\ref{tab:compositeness}. They are given for two different
models, the MCHM4 and MCHM5. Built in a five dimensional warped space, 
they provide a resummation of the full series in $\xi$, while the SILH
Lagrangian should be seen as an expansion in $\xi$ and can describe
composite Higgs models only in the vicinity of the SM limit. 
The bulk gauge symmetry $SO(5) \times U(1)_X \times SU(3)$ is broken down to
the SM gauge group on the ultraviolet boundary and to $SO(4) \times U(1)_X
\times SU(3)$ on the infrared. In the MCHM4 \cite{mchm4} the SM
fermions transform as spinorial representations, in the MCHM5 
\cite{mchm5} as fundamental representations of $SO(5)$. As a
consequence in the MCHM4 the Higgs couplings are changed universally as
function of $\xi$, while separately for SM vector bosons and fermions in
the MCHM5. The limits that can be obtained on the compositeness scale
$f$ range from below 1~TeV up to 5~TeV.
\begin{table}[hb!]
\begin{center}
\begin{tabular}{|l||c|c||c|c||c|}
\hline
$\xi$		& LHC	 & HL-LHC & LC	   & HL-LC & HL-LHC+HL-LC \\
\hline
universal	& 0.076  & 0.051  & 0.008  & 0.0052 & 0.0052	  \\
non-universal	& 0.068  & 0.015  & 0.0023 & 0.0019 & 0.0019	  \\
\hline \hline
$f$ [TeV]	&	 &	  &	   &	    &		  \\
\hline
universal	& 0.89	 & 1.09   & 2.82   & 3.41   & 3.41	  \\
non-universal	& 0.94	 & 1.98   & 5.13   & 5.65   & 5.65	  \\
\hline
\end{tabular}
\caption{Estimates of the parameter $\xi = (v/f)^2$ and the Goldstone
	     scale $f$ for various experimental set-ups and two
	     different fermion embeddings (universal/MCHM4,
             non-universal/MCHM5); from Ref.~\cite{coupreview}.}
\label{tab:compositeness}
\end{center}
\vspace*{-0.5cm}
\end{table}

\begin{figure}[hb]
\begin{center}
\includegraphics[width=0.49\textwidth]{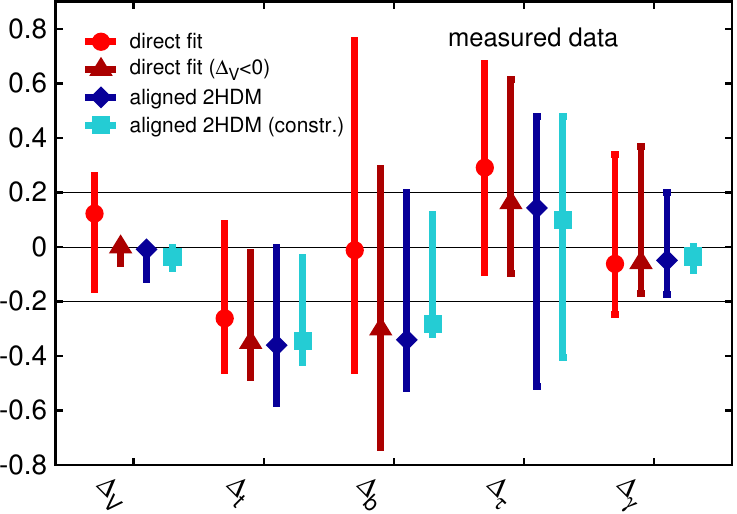}
\caption{For the Higgs coupling measurement based on all currently available
  ATLAS and CMS data ($\int {\cal L}= 4.6-5 (7 \mbox{ TeV})+12-21 (8
  \mbox{ TeV})$~fb$^{-1}$): Comparison of the fits to the
  weak-scale couplings with a fit to the aligned 2HDM in terms of
  the light Higgs couplings. Figure from Ref.~\cite{lopezval}.
}
\label{fig:2hdmcomp}
\end{center}
\end{figure}

As a last example, the interpretation of the current Higgs coupling
measurements in terms of a 2-Higgs-Doublet Model (2HDM) \cite{2hdm} is
shown. The Yukawa couplings of the two Higgs doublets are taken to
be proportional to each other in flavour space. At tree-level this
aligned 2HDM has five free parameters, where the mass of the charged
Higgs boson, which contributes to the effective Higgs-photon coupling,
is already included. For simplicity, custodial symmetry is assumed to
be fulfilled, {\it i.e.}~the deviations of the Higgs-gauge couplings
from the SM couplings are $\Delta_Z = \Delta_W \equiv \Delta_V
<0$. Figure~\ref{fig:2hdmcomp} shows the comparison of the extracted
free couplings according to Eq.~(\ref{eq:coupdev}) with the
corresponding fit to the aligned 2HDM parameters, translated into the
SM coupling deviations. The central values and the error bars agree
well between these two models. The observed small deviations are due
to correlations between the couplings induced in the 2HDM. If the
aligned 2HDM is realized in nature, additional constraints
arise from non-standard Higgs searches, from EWPTs and from flavour
constraints. They are taken into account in the cyan bands. For
recent work on Higgs coupling interpretations within the 2HDM with
respect to wrong-sign Yukawa couplings, see 
{\it e.g.}~\cite{Ferreira:2014naa}. 

\section{Conclusions}
After the discovery of the Higgs boson it is important to reveal the
true nature of the underlying dynamics of EWSB and to answer the question
if it is the Higgs boson of the SM or of some NP extension. In the
absence of any  discovery of new particles pointing to BSM physics, the Higgs
sector itself has to be explored in great detail and may turn out to be the
harbinger of NP. Composite Higgs models are examples of a Higgs
boson emerging from a strongly interacting sector. Although challenged
by EWPTs they are still a viable option. 
A wide class of BSM Higgs sectors can be studied in a rather
model-independent way through the effective Lagrangian approach. New
physics effects are encoded in higher 
dimensional operators that are built from SM fields and suppressed by
some high scale $\Lambda$ at which NP becomes effective. Higgs coupling
measurements prove useful to test such NP scales. In particular if
deviations in the Higgs couplings are due to mixings of the standard
Higgs with other new scalars, scales in the TeV range can be
constrained by the LHC, while coupling deviations due to loop effects
suffer from an additional loop suppression factor and are therefore
less sensitive to $\Lambda$. Coupling fits performed within specific 
models, finally, complement the interpretation within the effective Lagrangian
approach. With the increasing accuracy in the measurements at the next
run of the LHC new exciting physics may wait for us to be discovered. 

\Acknowledgements
I would like to thank the organisers of LHCP14 for the invitation to
give this talk, for the perfectly organised conference and for the pleasant
atmosphere with a lot of inspiring discussions. I greatly acknowledge
discussions with C.~Grojean, S.~Gupta, A.~Pomarol, M.~Rauch, R.~Santos
and M.Spira. 


\end{document}

%% file: econfmacros.tex
\def\beq{\begin{equation}}
\def\eeq#1{\label{#1}\end{equation}}
\def\eeqn{\end{equation}}

\def\beqa{\begin{eqnarray}}
\def\eeqa#1{\label{#1}\end{eqnarray}}
\def\eeqan{\end{eqnarray}}

\let\bar=\overbar

\def\Dslash{\not{\hbox{\kern-4pt $D$}}}
\def\dslash{\not{\hbox{\kern-2pt $\del$}}}

\def\msb{{\bar{\ssstyle M \kern -1pt S}}}

